\documentstyle[12 pt]{article}

\begin{document}

\ \vskip 0.5 in

\begin{center}
 { \Large {\bf String Theory, Quantum Mechanics}}
{\Large {\bf  and Noncommutative Geometry}}

\smallskip

  {\it - A new perspective on the gravitational dynamics of D0-Branes - } 

\bigskip


\bigskip

\bigskip

{{\large
{\bf T. P. Singh\footnote{e-mail address: tpsingh@tifr.res.in} }}}

\medskip

{\it Tata Institute of Fundamental Research,}\\
{\it Homi Bhabha Road, Mumbai 400 005, India.}
\vskip 0.5cm
\end{center}

\vskip 0.5 in

\begin{abstract}

\noindent We do not know the symmetries underlying string theory. 
Furthermore, there must exist an inherently quantum, and spacetime independent,
formulation of this theory. Independent of string theory, there should
exist a description of quantum mechanics which does not refer to  a classical
spacetime manifold. We propose such a formulation of quantum mechanics, based
on noncommutative geometry. This description reduces to standard quantum
mechanics, whenever an external classical spacetime is available. However,
near the Planck energy scale, self-gravity effects modify the Schr\"odinger
equation to the non-linear Doebner-Goldin equation. Remarkably, this 
non-linear equation also arises in the quantum dynamics of D0-branes.
This suggests that the noncommutative quantum dynamics proposed here is
actually the quantum gravitational dynamics of D0-branes, and that automorphism
invariance is a symmetry of string theory.

\end{abstract}

\vskip 1.0 in

\centerline{\it This essay received an honorable mention in the}
\centerline{\it Gravity Research Foundation Essay Competition, 2006}

\newpage

\noindent String theory has had numerous successes. However, one still does 
not know what are the symmetry principles which underlie the 
theory \cite{witten}. One can get a glimpse of the underlying symmetry by
recalling two facts. Firstly, spacetime is an emergent concept in string 
theory; hence there must exist a more complete formulation of the theory 
which does not refer to a classical spacetime. Secondly, duality symmetries
can map a strongly quantum process in one string theory to a weakly quantum
process in another string theory, suggesting that the fundamental formulation
of string theory/M-theory should be inherently quantum, and not merely the 
result of a traditional quantization of a classical theory. Together, these 
two facts show that there should exist an `inherently quantum' formulation of 
string theory/M-theory, which should not refer to a classical spacetime.

Independent of string theory, there must exist a more fundamental
formulation of quantum mechanics, which does not refer to a classical
spacetime. This is because time evolution can naturally be defined only
when a spacetime manifold is given. Now, the spacetime metric which overlies 
the classical manifold is produced by classical matter fields. In 
principle the Universe could be in a state in which there are no classical
matter fields. The spacetime metric would then undergo quantum fluctuations,
and in accordance with the Einstein hole argument \cite{joy} one can no
longer assign any physical significance to the underlying classical
spacetime manifold. In this essay we shall propose such a fundamental
formulation of quantum mechanics (without any prior motivation from string 
theory) using the language of noncommutative differential geometry. In the 
process we find that the new formulation resembles a description of the 
gravitational dynamics of D0-branes in string theory/M-theory. This suggests 
that the sought for underlying symmetry of string theory/M-theory may have a 
deep connection with the principles of noncommutative geometry.

Again, independent of string theory, there is additional evidence that
Einstein equations should not be quantized in the traditional mold, because
they arise as a thermodynamic equation of state \cite{jp} when one assumes a 
proportionality between horizon area and entropy, along with the Clausius
relation $dQ=TdS$, for all local Rindler horizons in a spacetime. Classical 
gravitation is seen as an emergent phenomenon, and hence so is the presently 
known formulation of 
quantum mechanics, which requires an external classical gravitational field 
to be specified, so as to define time evolution. 
Noncommutative geometry provides an arena for a theory of quantum 
gravity, from which both general relativity and quantum mechanics emerge as 
approximations. 

The topological and differential properties of a differentiable Riemannean 
manifold can be expressed algebraically in terms of the so-called spectral
triple, which consists of a commutative algebra ${\cal A}$, a Hilbert space  
${\cal H}$ and a self-adjoint operator ${\cal D}$. When the algebra is made
noncommutative, the geometry that is correspondingly obtained is called
noncommutative differential geometry - it is a  natural
generalization of Riemannean geometry. Diffeomorphisms on the manifold can
be mapped to the automorphisms of the algebra; when the algebra is 
noncommutative its automorphisms represent corresponding diffeomorphisms 
on the noncommutative space. If a physical  noncommutative space possesses a
symmetry representing invariance under diffeomorphisms, this symmetry can be 
expressed as an invariance under the algebra automorphisms - we  call
such a symmetry automorphism invariance. An attractive property of
a noncommutative geometry is the Tomita-Takesaki theorem, which defines
a one-parameter group of automorphisms of the algebra, hence giving the
possibility of defining the much needed concept of time in a quantum 
theory of gravity \cite{conrov}.  
    
We propose that noncommutative geometry is the 
framework for a formulation of quantum mechanics which does
not refer to a classical (i.e. Riemannean) spacetime manifold. We suggest the
concept of a noncommuting coordinate system, which `covers' a noncommutative
manifold, wherein commutation relations between coordinates are 
introduced on physical grounds. The new formulation of quantum dynamics,
which is given in such a coordinate system, is invariant under
transformation (automorphism) from the given coordinate system to another
noncommuting coordinate system. This generalizes general covariance to the
noncommutative case, and we are proposing that the theory is invariant under
transformations of noncommuting coordinates. This formulation satisfies two
important properties. In the limit in which the system becomes macroscopic,
the noncommutative spacetime is indistinguishable from  ordinary
commutative spacetime, and the dynamics reduces to classical dynamics. 
Secondly, if a dominant part of the system becomes macroscopic and classical,
and a sub-dominant part remains quantum (as our Universe is) then seen from 
the viewpoint of the dominant part, the quantum dynamics of the sub-dominant
part is the same as the quantum dynamics known to us. In this sense, standard
quantum mechanics is recovered from its underlying noncommutative formulation.

Consider a system of quantum mechanical particles having a total mass-energy
much less than Planck mass $m_{Pl}$, and assume that no external classical 
spacetime manifold is available. Since Planck mass scales inversely with the
gravitational constant, we are justified here in neglecting the gravitational
field, and the resulting quantum spacetime produced by the system will be
called a `quantum Minkowski spacetime'. To describe the dynamics using
noncommutative geometry consider a particle with mass $m\ll m_{Pl}$    
in a 2-d noncommutative spacetime with coordinates ($\hat{x}, \hat{t})$.
Generalization to higher dimensions and the multi-particle case is
straightforward \cite{singh}. (Later below we will consider the important case
where the mass-energy of the system is comparable to Planck mass and gravity
can no longer be neglected).

On the quantum Minkowski spacetime we introduce the non-Hermitean 
flat metric 
\begin{eqnarray}
\label{ncfm}  
\hat{\eta}_{\mu\nu} = \left(\begin{array}{cc}
                      1 & 1 \\
                      -1 & -1 \end{array} \right)
\end{eqnarray}
and the corresponding noncommutative line-element
\begin{equation}
ds^{2}=\hat{\eta}_{\mu\nu}d\hat{x}^{\mu}d\hat{x}^{\nu}=
d\hat{t}^{2}-d\hat{x}^{2}
+d\hat{t}d\hat{x}-d\hat{x}d\hat{t}
\label{lin}
\end{equation}
which is invariant under a generalized Lorentz transformation 
\cite{singh}.

Noncommutative dynamics is constructed by defining a velocity  $\hat{u}^{i}=
d\hat{x}^{i}/ds$,
which, from (\ref{lin}), satisfies the relation
\begin{equation}
1=\hat{\eta}_{\mu\nu}\frac{d\hat{x}^{\mu}}{ds}\frac{d\hat{x}^{\nu}}{ds}=
(\hat{u}^{t})^{2}-(\hat{u}^{x})^{2} + 
\hat{u}^{t}\hat{u}^{x} - \hat{u}^{x}\hat{u}^{t}  
\label{vel}
\end{equation} 
and by defining the generalized momentum as 
$\hat{p}^{i}=m\hat{u}^{i}$, 
which hence satisfies 
\begin{equation}
\hat{p}^{\mu}\hat{p}_{\mu} = m^{2}.
\label{nchj}
\end{equation}
Here, $\hat{p}_{\mu}=\hat{\eta}_{\mu\nu}\hat{p}^{\mu}$ is well-defined. Written
explicitly, this equation becomes
\begin{equation}
(\hat{p}^{t})^{2}-(\hat{p}^{x})^{2} + 
\hat{p}^{t}\hat{p}^{x} - \hat{p}^{x}\hat{p}^{t}  = m^{2}.
\label{nce}
\end{equation} 
Dynamics is constructed by introducing a {\it complex} action 
$S(\hat{x},\hat{t})$ and by defining the momenta introduced above as
gradients of this complex action. In analogy with classical mechanics this
converts (\ref{nce}) into a (noncommutative) Hamilton-Jacobi equation,
which describes the dynamics.

When an external classical Universe with a classical manifold
$(x,t)$ becomes available (see below), one defines the generalized momentum 
$(p^{t},p^{x})$ in terms of the complex action $S(x,t)$ as 
\begin{equation}
p^{t}=-{\partial S\over \partial t}, \qquad p^{x} = 
{\partial S \over \partial x}
\label{pmoo}
\end{equation}
and from (\ref{nce}) the following fundamental rule for relating 
noncommutative dynamics to standard quantum dynamics \cite{singh}
\begin{equation}
(\hat{p}^{t})^{2}-(\hat{p}^{x})^{2} + 
\hat{p}^{t}\hat{p}^{x} - \hat{p}^{x}\hat{p}^{t}  = ({p}^{t})^{2}-({p}^{x})^{2}
 + i\hbar {\partial p^{\mu}\over \partial x^{\mu}}=m^{2}.
\label{nceq}
\end{equation}
In terms of the complex action the right hand side of this equation can be
written as
\begin{equation}
\left({\partial{S}\over \partial t}\right)^{2}-\left({\partial{S}\over \partial x}\right)^{2}
-i\hbar\left({\partial^{2}S\over\partial t^{2}}-{\partial^{2}S\over\partial x^{2}}\right)=m^{2}\label{hjc}
\end{equation}
and from here, by defining a quantum state $\psi$ in a natural manner:
 $\psi=e^{iS/\hbar}$, we arrive at the Klein-Gordon equation 
 \begin{equation}-\hbar^{2}
\left({\partial^{2}\over\partial t^{2}}-{\partial^{2}\over\partial x^{2}}\right)\psi=m^{2}\psi\label{kg},
\end{equation}
as desired.

The commutation relations on the non-commutative spacetime are \cite{singh}
\begin{equation}
[\hat{t},\hat{x}]=iL_{Pl}^{2}, \qquad [\hat{p}^{t}, \hat{p}^{x}]=iP_{Pl}^{2}.
\label{commu}
\end{equation}
It is the momentum space noncommutativity which makes our formulation
fundamentally different from previous applications of noncommutative
geometry to quantum gravity.  On the noncommutative 4-d `phase space' corresponding to the
two noncommuting coordinates $(\hat{x},\hat{t})$ there is a natural grid 
(which is a consequence of noncommutativity) which when projected on the
$(\hat{x},\hat{p}^{x})$ plane, has an area $L_{Pl}\times P_{PL}=\hbar$. When we
examine this noncommutative quantum dynamics from our ordinary 
spacetime, we take the limit $L_{Pl}\rightarrow 0, P_{Pl} \rightarrow \infty$,
while keeping their product (the area of the fundamental phase space cell)
constant at $\hbar$. From the viewpoint of our classical spacetime, quantum 
dynamics is then recovered by imposing the commutator $[q,p]=i\hbar$, which 
preserves the granular structure of the phase space, in quantum mechanics. 

If the mass-energy of the particle is not negligible
in comparison to Planck mass its self-gravity must be taken into 
account. The metric (\ref{ncfm}) gets modified to
\begin{eqnarray}
\label{nccm}  
\hat{h}_{\mu\nu} = \left(\begin{array}{cc}
                      \hat{g}_{tt} & \hat{\theta} \\
                      -\hat{\theta} & -\hat{g}_{xx} \end{array} \right)
\end{eqnarray}
and Eqns.(\ref{lin}), (\ref{nchj}), (\ref{nce}), (\ref{nceq}) and 
(\ref{hjc}) are respectively replaced by the equations
\begin{equation}
ds^{2}=\hat{h}_{\mu\nu}d\hat{x}^{\mu}d\hat{x}^{\nu}=
\hat{g}_{tt}d\hat{t}^{2}-\hat{g}_{xx}d\hat{x}^{2}
+\hat\theta[d\hat{t}d\hat{x}-d\hat{x}d\hat{t}],
\label{linc}
\end{equation}
\begin{equation}
\hat{h}_{\mu\nu}\hat{p}^{\mu}\hat{p}^{\nu}=m^{2}, 
\end{equation}
\begin{equation}
\label{nceq2}\hat g_{tt}(\hat p^t)^2-\hat g_{xx}(\hat p^x)^2+\hat \theta
\left( \hat p^t\hat p^x-\hat p^x\hat p^t\right) =m^2,
\end{equation}
\begin{equation}
\label{corr}\hat g_{tt}(\hat p^t)^2-\hat g_{xx}(\hat p^x)^2+\hat \theta
\left( \hat p^t\hat p^x-\hat p^x\hat p^t\right)=g_{tt}({p}^t)^2-g_{xx}({p}%
^x)^2+i\hbar \theta {\frac{\partial p^\mu }{\partial x^\mu }},
\end{equation}
\begin{equation}
\label{nceq3}g_{tt}({p}^t)^2-g_{xx}({p}%
^x)^2+i\hbar \theta {\frac{\partial p^\mu }{\partial x^\mu }}=m^2.
\end{equation}
Like in general relativity, the metric $\hat{h}_{\mu\nu}$ is 
determined by the mass $m$ via the quantum state $S(\hat{x},\hat{t})$. The
field equations are covariant under general coordinate transformations
(automorphisms) of noncommuting coordinates. In the macroscopic limit 
$m\gg m_{Pl}$ the antisymmetric component $\theta$ goes to zero; the 
noncommutative spacetime (\ref{linc}) is indistinguishable from ordinary 
commutative spacetime, and Eqn. (\ref{nceq3}) reduces to classical dynamics.   

We are now ready to make contact with string theory. If we substitute
for the momenta in (\ref{nceq3}) in terms of the complex action using
(\ref{pmoo}) and then substitute $\psi=e^{iS/\hbar}$ and take the
non-relativistic limit, the resulting effective Schr\"odinger equation is
non-linear \cite{singh}. It is the same as the Doebner-Goldin equation
\cite{dg} - the latter arises when one classifies physically different
 quantum systems by considering unitary representations of the group of
diffeomorphisms $Diff(R^{3})$. The simplest case is obtained when in 
(\ref{nceq3}) one approximates the diagonal metric components to unity,
giving the non-linear Schr\"odinger equation 
\begin{equation}
i\hbar\frac{\partial\psi}{\partial t} = -\frac{\hbar^{2}}{2m}\frac
{\partial^{2}\psi}{\partial x^{2}} + \frac{\hbar^{2}}{2m}(1-\theta)
\left(\frac{\partial^{2}\psi}{\partial x^{2}} - 
\left[\frac{\partial (\ln\psi)}{\partial x}\right]^{2}\psi
\right).
\label{nlse}
\end{equation}

In string theory, it is known that the coordinates of D0-branes, being
matrices, satisfy a noncommutative geometry. When one considers the quantum
dynamics of D0-branes, an effective non-linear Schr\"odinger equation is 
arrived at, because of the quantum recoil due to the exchange of
string states between individual D-particles \cite{mav}. 
This is analogous to considering the effect of self-gravity on the dynamics, 
and remarkably, this non-linear equation is again the Doebner-Goldin equation! 
Now, this equation
has a very specific non-linearity structure, and it could not be a 
coincidence that it should arise in two different contexts, both
concerned with quantum spacetime. We hence propose that automorphism
invariance, in the manner introduced here, is one of the fundamental
symmetries of string theory, and the noncommutative dynamics 
described above is actually the quantum gravitational dynamics of 
D0-branes. It is inherently quantum mechanical and does not refer to  a 
classical spacetime, as desired in string theory.

\end{document}